\documentclass[11pt,twoside]{article}
\usepackage{asp2010}

\resetcounters

\begin{document}

\title{Heavy Element Abundance Patterns in Hot White Dwarfs from a Survey of the FUSE Archive}
\author{M.A. Barstow,$^1$, S. L. Casewell$^1$, J.B. Holberg$^2$ and J.K. Barstow$^3$}
\affil{$^1$Department of Physics \& \ Astronomy, University of Leicester, University Road, Leicester LE1 7RH, UK}
\affil{$^2$Lunar and Planetary Laboratory, University of Arizona, Tucson, AZ, USA}
\affil{$^3$Department of Astrophysics, University of Oxford, Denys Wilkinson Building, Keble Road, Oxford, OX1 3RH}

\begin{abstract}
We present a series of systematic abundance measurements for 89 hot DA white dwarfs drawn from the FUSE observation archive. These stars span the temperature range $\approx$ 20000-70000K, and form the largest sample to-date, exceeding our earlier study, based mainly on IUE and HST data, by a factor three. Using the heavy element blanketed non-LTE stellar atmosphere calculations from this previous work, we are able to measure the abundances of carbon, silicon, phosphorus and sulphur and examine how they change as the stars cool. We are able to establish the broad range of abundances seen in a given temperature range and establish the incidence of stars which, like HZ43, appear to (surprisingly) be completely devoid of any material other than H in their atmospheres. As a result we can begin to identify stars with peculiar abundances in this temperature range and determine whether or not these objects might be accreting planetary debris, as has been inferred for many cooler objects.
\end{abstract}

\section{Introduction}
Measurements of the physical parameters of white dwarf stars are crucial to understanding their evolution and their relationship with progenitor stars on the giant branch. Of particular importance are accurate measurements of photospheric composition and structure, which can tell us how the He and heavy element content of white dwarfs may change as the stars cool. 

Our understanding of the composition of hot white dwarfs has evolved radically during the past decades. \citet{schatzman58} pointed out that the strong gravitational field (log g $\approx$ 8 at the surface) of a white dwarf would cause rapid downward diffusion of elements heavier than the principal H or He components, yielding very pure atmospheres devoid of any other material. However, many white dwarfs show evidence for significant quantities of heavy elements in their atmospheres (e.g. \citealt{barstow93, marsh97, barstow03}), which appear to be present because radiation pressure (or "radiative levitation") counteracts the downward force of gravity (e.g. \citealt{vauclair79, chayer95, schuh02}).

In the last few years an increasing number of heavy element-contaminated white dwarfs have been discovered, associated with infrared excesses and closely orbiting rings of dusty and gaseous debris \citep{farihi09, jura09, gaensicke08}. Such debris offers a ready source of material for accretion into the atmospheres of the host stars. A statistical analysis of the galactic position and kinematics of DZ and DZA stars from the Sloan Digital Sky Surveys reported no correlation between accreted Ca abundances and spatial--kinematical distributions relative to interstellar material \citep{farihi10}. Furthermore, they find that the DZ and DC stars belong to the same population. Accretion from the ISM cannot simultaneously account for the polluted and non-polluted sub-populations. Therefore, it is probable that circumstellar matter, the rocky remnants of terrestrial planetary systems, contaminates these white dwarfs.

\citet{barstow03} published a survey of hot DA white dwarf compositions based on far-UV spectra obtained with IUE and HST, the largest to date. However, the 25 stars included in this work provide a sample with a number of in-built biases. For example, the spectra were mostly obtained where it was expected that there was a high chance of detecting heavy elements. As a result, most of the targets observed have T$_{\rm eff}$ $\approx$ 50000K or above, with very sparse coverage at lower temperatures. In addition, it is not possible to say anything about the frequency of stars with or without heavy elements.

The FUSE mission observed a large number ($>$100) hot white dwarf stars during its lifetime. As these targets were derived from a variety of programmes, including studies of the ISM as well as white dwarf composition, the coverage of the white dwarf temperature range is much more uniform and the selection biases less problematic. Therefore, a detailed analysis of this sample can provide insight into the photospheric abundance patterns across the whole temperature range and the proportion of stars which are uncontaminated by photospheric heavy elements.

\section{The FUSE Sample}

While more than 100 hot white dwarfs and related objects were observed during the course of the FUSE mission, only 89 of these have spectra with sufficient signal to noise for the the purposes of this study. Nevertheless, this is a factor three increase over the earlier study of \citet{barstow03}. The wavelength coverage of 900-1180\AA \ is complementary to that of IUE and HST ($\approx$ 1100-1700\AA), but has a limited range of overlap. Therefore, analysis of the FUSE data relies on a different group of atoms and ions to those available at longer wavelengths. The principle species detected in the FUSE wavelength range are: C\sc{iii}, C\sc{iv}, N\sc{iv}, O\sc{vi}, Si\sc{iv}, P\sc{iv}, P\sc{v}, S\sc{iv} \rm and S\sc{vi}\rm. However, not all of these are useful for abundance measurements due to ISM contamination (O\sc{vi})\rm, or the signal to noise in that region of the spectrum (N\sc{iv} \rm and S\sc{vi}\rm), and lack of sensitivity to abundance changes in this temperature range (P\sc{iv})\rm. Hence, the principal useful lines are the C\sc{iii} \rm multiplet near 1175\AA , Si\sc{iv} 1066.614/1122.5\AA , P\sc{v} 1117.977/1128.008\AA \ \rm and S\sc{iv} \rm 1062/1072.793\AA . 

\section{Results}

\begin{figure}
\includegraphics[scale=0.5]{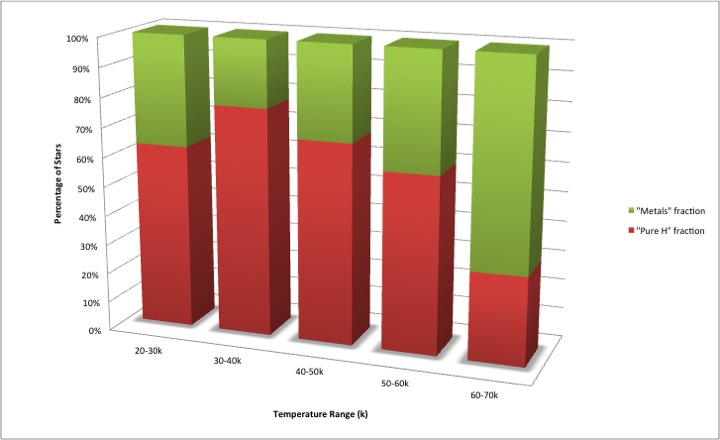}
\caption{Histogram showing the fraction of DA white dwarfs with (green) and without (red) photospheric heavy elements in 10 000K effective temperature bins.}
\label{fig1}
\end{figure}

We will report the detailed abundance measurements for each star in a forthcoming journal paper (Barstow et al. 2012, in preparation). Here, we summarise the overall results for the complete sample of stars. Table \ref{table1} lists the stars included in the sample, along with their effective temperatures and log g values. For each star, detection is indicated by  "D" and non-detection by "U". It is clear from studying the whole sample, that in a particular temperature range there are numbers of stars both with and without photospheric heavy elements. We define a metal-containing star as one which has a detection of at least one of the species listed above, but many objects contain more than one detected element. Figure \ref{fig1} illustrates this, where the whole temperature range is divided into 10000K bins and the relative proportions of those with metals and those with pure H atmospheres are represented by the histogram. The relative fractions are not uniform and the highest proportion of "metal-rich" DAs is found in the highest temperature range, as might be expected. An important result is that, even at the highest temperatures, there is a significant proportion of white dwarfs that appear, at the limit of the sensitivity of the observations, to have pure H envelopes. Furthermore, a significant fraction of stars have metal-containing atmospheres even at the lowest temperatures.

\begin{table}\footnotesize

\begin{center}
\caption{\label{table1}The stars included in the sample, along with their effective temperatures and log g values. For each star, detection is indicate by  "D" and non-detection by "U".}
\begin{tabular}{l c c c c c c | l c c c c c c}
\hline
Name& T$_{\rm eff}$ & log g& C& Si& P& S &Name& T$_{\rm eff}$ & log g& C& Si& P& S\\
& K&&&&&&&K&&&&&\\
\hline

WD2350-706 & 69300 & 8.00 & U & D & D & D& WD1057+719 & 39555 & 7.66 & U & U & U & U\\
WD1056+516 & 68640 & 8.08 & U & U & U & U& WD2111+498 & 38866 & 7.84 & U & D & D & D\\
WD2146-433 & 67912 & 7.58 & U & D & D & D& WD1611-084 & 38500 & 7.85 & D & D & D & U\\
WD1342+442 & 66750 & 7.93 & D & D & D & D& WD1254+223 & 38205 & 7.90 & U & U & U & U\\
WD2211-495 & 65600 & 7.42 & U & D & D & D& WD1615-154 & 38205 & 7.90 & U & U & U & U\\
WD0229-481 & 63400 & 7.43 & U & D & D & D& WD2257-073 & 38010 & 7.84 & U & U & U & U\\
WD0232+035 & 62947 & 7.53 & U & D & D & D& WD1648+407 & 37850 & 7.95 & U & U & U & U\\
WD0621-376 & 62280 & 7.22 & D & D & D & D& WD1845+683 & 36888 & 8.12 & U & U & U & U\\
WD1711+668 & 60900 & 8.39 & U & U & U & U& WD1109-225 & 36750 & 7.50 & U & U & U & U\\
WD0027-636 & 60595 & 7.97 & U & U & U & U& WD1603+432 & 36257 & 7.85 & U & U & U & U\\
WD0455-282 & 58080 & 7.90 & U & D & D & D& WD1636+351 & 36056 & 7.71 & U & U & U & U\\
WD0501+524 & 57340 & 7.48 & D & D & D & D& WD0937+505 & 35552 & 7.76 & U & U & U & U\\
WD2331-475 & 56682 & 7.64 & U & D & D & D& WD1021+266 & 35432 & 7.48 & D & D & D & D\\
WD1234+481 & 55570 & 7.57 & U & U & U & U& WD0416+402 & 35227 & 7.75 & U & U & U & U\\
WD1725+586 & 54550 & 8.49 & U & U & U & U& WD2124+191 & 35000 & 9.00 & U & U & U & U\\
WD2116+736 & 54486 & 7.76 & U & U & U & U& WD0050-332 & 34684 & 7.89 & U & U & D & D\\
WD0354-368 & 53000 & 8.00 & U & U & U & U& WD0236+498 & 33822 & 8.47 & U & U & U & U\\
WD1921-566 & 52946 & 8.16 & U & U & U & U& WD1942+499 & 33500 & 7.86 & U & D & D & U\\
WD2309+105 & 51300 & 7.91 & U & D & D & D& WD0603-483 & 33040 & 7.80 & U & U & U & U\\
WD0226-615 & 50000 & 8.15 & U & U & U & U& WD1917+509 & 33000 & 7.90 & U & D & D & U\\
WD1314+293 & 49435 & 7.95 & U & U & U & U& WD0353+284 & 32984 & 7.87 & U & U & U & U\\
WD2124-224 & 48297 & 7.69 & U & D & D & U& WD0320-539 & 32860 & 7.66 & U & U & U & U\\
WD0004+330 & 47936 & 7.77 & U & U & U & U& WD0549+158 & 32780 & 7.83 & U & U & D & U\\
WD1040+492 & 47560 & 7.62 & U & U & U & U& WD0235-125 & 32306 & 8.44 & U & U & U & U\\
WD2011+398 & 47057 & 7.74 & D & D & D & D& WD1844-223 & 31470 & 8.17 & U & U & U & U\\
WD1528+487 & 46230 & 7.70 & U & U & U & U& WD0809-728 & 30585 & 7.90 & U & U & U & U\\
WD0001+433 & 46205 & 8.85 & U & U & U & U& WD1620+647 & 30184 & 7.72 & U & U & U & U\\
WD2321-549 & 45860 & 7.73 & U & D & D & D& WD0147+674 & 30120 & 7.70 & U & U & U & U\\
WD2152-548 & 45800 & 7.78 & U & U & U & U& WD0830-535 & 29330 & 7.79 & U & U & U & U\\
WD0802+413 & 45394 & 7.39 & U & U & U & U& WD1019-141 & 29330 & 7.79 & D & U & D & D\\
WD1819+580 & 45330 & 7.73 & D & D & D & D& WD0252-055 & 29120 & 7.50 & U & D & D & U\\
WD1029+537 & 44980 & 7.68 & U & U & U & U& WD1041+580 & 29016 & 7.79 & U & U & U & U\\
WD0131-164 & 44850 & 7.96 & U & D & D & U& WD1734+742 & 28795 & 8.00 & D & D & U & U\\
WD1631+781 & 44559 & 7.79 & U & U & U & U& WD2020-425 & 28597 & 8.54 & U & U & U & U\\
WD2000-561 & 44456 & 7.54 & D & D & D & D& WD0106-358 & 28580 & 7.90 & D & D & D & U\\
WD0715-704 & 44300 & 7.69 & U & U & U & U& WD2014-575 & 26579 & 7.78 & U & U & U & U\\
WD2004-605 & 44200 & 8.14 & U & U & U & U& WD2043-635 & 25971 & 8.36 & U & U & U & U\\
WD1800+685 & 43701 & 7.80 & U & U & U & U& WD0457-103 & 25540 & 8.20 & U & U & U & U\\
WD1440+753 & 42400 & 8.54 & U & U & U & U& WD0905-724 & 25398 & 7.35 & U & U & U & U\\
WD0346-011 & 42373 & 9.00 & U & U & U & U& WD0041-092 & 24900 & 7.50 & U & U & U & U\\
WD1024+326 & 41354 & 7.59 & U & U & U & U& WD0512+326 & 22750 & 8.01 & U & D & U & U\\
WD1950-432 & 41339 & 7.85 & U & D & U & U& WD1337+701 & 20435 & 7.87 & D & D & D & U\\
WD0659+130 & 39960 & 8.31 & U & U & U & U& WD1635+529 & 20027 & 8.14 & U & U & U & U\\
WD1302+597 & 39960 & 8.31 & U & U & U & U& WD0310-688 & 16181 & 8.06 & U & U & U & U\\
WD1550+130 & 39910 & 6.82 & U & U & U & U&            &       &      &   &   &   &  \\
\hline

\end{tabular}
\end{center}
\end{table}

\begin{figure}
\includegraphics[scale=0.5]{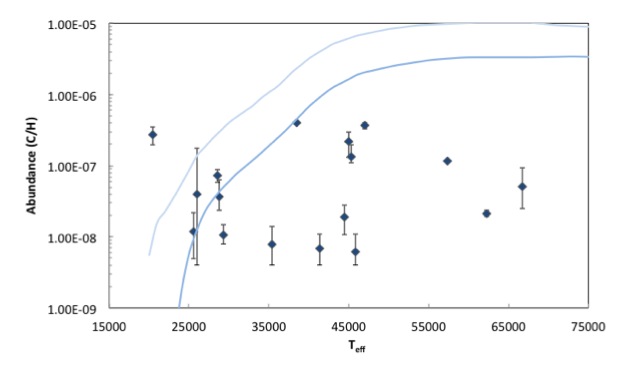}
\caption{Measured abundance of carbon (filled diamonds) as a function of effective temperature compared to the predictions of radiative levitation calculations for log g=7.5 (thin curve) and log g=8.0 (thick curve).}
\label{fig2}
\end{figure}

\begin{figure}
\includegraphics[scale=0.5]{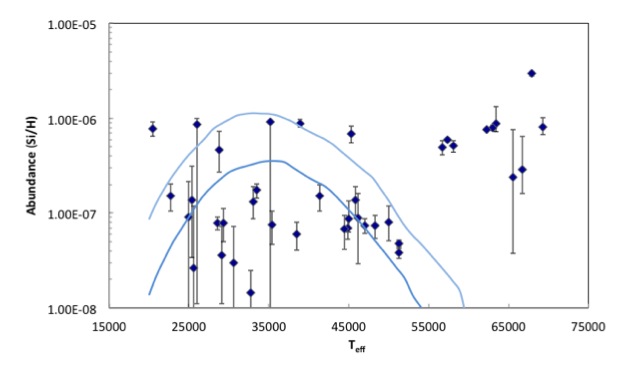}
\caption{Measured abundance of silicon (filled diamonds) as a function of effective temperature compared to the predictions of radiative levitation calculations for log g=7.5 (thin curve) and log g=8.0 (thick curve).}
\label{fig3}
\end{figure}

Figures \ref{fig2} and \ref{fig3} show the measured abundance of C and Si as a function of effective temperature. The filled diamonds are the measurements with their associated error bars while the curves represent the abundances predicted by the radiative levitation theory of \citet{chayer95} for log surface gravities of 7.5 (thin curve) and 8.0 (thick curve) spanning the range of log g for the sample of stars. It is clear that where C is detected the predicted abundances are mostly significantly higher than observed, except at temperatures below $\approx$ 27000K. In contrast, there is better agreement between the predictions and measured Si abundances over much of the temperature range, although at the highest temperatures (above $\approx$ 55000K) the measured values are several orders of magnitude above what is expected. At intermediate temperatures, in the range 25000-40000K, observed abundances are somewhat lower than predicted. Similar patterns are also seen for P and S (not shown here due to limited space). One significant features of all the measurements is the large scatter in values at any particular temperature. Much larger than would be expected just from gravity variations between objects.

\section{Discussion}

We have reported on the largest survey of hot DA white dwarf compositions carried out to-date. The most important outcome is that both metal-containing and pure H atmospheres exist at all temperatures in the range covered by the study. It is also clear that the measured abundance values and their patterns bear little relationship to the abundances predicted by radiative levitation calculations. This indicates that these are incomplete in some way and should be revisited.

\acknowledgements MAB and JKB acknowledge the support of the Science and Technology Facilities Council. SLC is supported by the University of Leicester. JBH acknowledges a visiting professorship from the University of Leicester.

\bibliography{barstow_paper}

\end{document}